\documentclass[showpacs,amsmath,amssymb,showkeys,floatfix]{revtex4}
\usepackage{amsfonts,amsmath}
\usepackage{graphicx}
\usepackage{dcolumn}
\usepackage{bm}
\usepackage{url}

\DeclareFontFamily{OT1}{pzc}{}
\DeclareFontShape{OT1}{pzc}{m}{it}{<-> s * [1.10] pzcmi7t}{}
\DeclareMathAlphabet{\mathpzc}{OT1}{pzc}{m}{it}

\newcommand{\qslash}{\not{\hbox{\kern-2.3pt $q$}}}
\newcommand{\kslash}{\not{\hbox{\kern-2.3pt $k$}}}
\newcommand{\pslash}{\not{\hbox{\kern-2.3pt $p$}}}

\newcommand{\Pslash}{\not{\hbox{\kern-2.3pt $P$}}}
\newcommand{\Pslashsup}{^\not{\hbox{\kern-0.5pt $^P$}}}

\begin{document}
\title{Optical dispersion of composite particles consisting of millicharged constituents}

\author {Audrey~K.~Kvam}
 
\author{David C.~Latimer}

\affiliation{Department of Physics, University of Puget Sound,
Tacoma, WA 98416-1031 
}

\begin{abstract}

Composite dark matter (DM) comprised of electrically charged constituents can interact with the electromagnetic field via the particle's dipole moment.  This interaction results in a dispersive optical index of refraction for the DM medium.  We compute this refractive index for atomic dark matter and more strongly bound systems, modeled via a harmonic oscillator potential.  The dispersive nature of the index will result in a time lag between high and low energy photons simultaneously emitted from a distant astrophysical observable.  This time lag, due to matter dispersion, could confound potential claims of Lorentz invariance violation (LIV) which can also result in such time lags.  We compare the relative size of the two effects and determine that  the dispersion due to DM is dwarfed by potential LIV effects for energies below the Planck scale.

\end{abstract}

\maketitle

\section{Introduction}

A concordance of observations support a universe whose energy budget is dominated by the unknown elements of  dark energy (DE) and dark matter (DM), while baryonic matter occupies only around 5\% of the total energy \cite{concord1,concord2,wmap9,planck2013,planck2015}.  Focusing on DM, all concrete evidence for its existence is solely based upon its gravitational interactions, prompting some to hypothesize alternative explanations to DM, like modified gravitational interactions (e.g., Ref~\cite{mond}), but these alternatives are not viable in light of  observations of colliding galaxy clusters where it is shown that the bulk of the clusters consists of non-luminous matter that does not interact (except gravitationally) \cite{bullet}.
 Though there is no evidence that DM interacts through non-gravitational channels,  theories beyond the standard model that incorporate DM candidates often contain a rich panoply of interactions between  dark and standard model particles.  In fact, DM will interact electromagnetically even if it is electrically neutral, provided that it couples to charged particles.  Granted, these electromagnetic interactions are suppressed,  but they are generally nonzero.  Given this, a DM medium will have an optical index of refraction which is generally dispersive; i.e., the phase velocity of light in the medium is frequency dependent.

In Ref.~\cite{dm_n}, we computed the refractive index for various particulate DM models. The forward Compton scattering amplitude  links  the medium's optical properties with the particle-level interaction between the photon and DM \cite{goldwatson,fermi}. Rather general considerations (namely, Lorentz covariance and invariance under charge conjugation, parity, and time-reversal symmetries) restrict the structure of this forward Compton amplitude at low photon energies, $\omega$.  
As a result, the leading order contributions to the forward scattering amplitude are model independent, attributable to the charge, mass, and magnetic dipole moment of the scatterer \cite{GGT,low,GG,G,lapidus}, and the higher order, model dependent, terms follow a known form such that the index is $n(\omega) = 1 - A\omega^{-2} +B +C \omega^2 +\dots$ with each coefficient non-negative assuming small $\omega$ \cite{cosmicn, dm_n}.

For electrically neutral DM candidates, the $\mathcal{O}(\omega^{-2})$ term in the index of refraction vanishes because $A = 2 q^2$ where $q$ is the electric charge of the DM.  The resulting index of refraction simplifies to $n(\omega) \approx 1 + B + C \omega^2$.  In principle, one can experimentally assess the coefficient $C$ through astrophysical observation.  Given the normally dispersive nature of the DM in the cosmos,  high energy photons will travel more slowly than ones with lower energy.  If a broadband  pulse of  photons travels over a sufficient baseline through the DM medium, then, statistically, the arrival time of photons from that pulse will be energy dependent.  For this study, a near ideal source of photons is a  gamma ray burst (GRB), observable out to redshifts of $z>9$ \cite{Cucchiara:2011pj}. So, if the arrival time of photons from a large sample of GRBs shows energy and baseline dependence characteristic of matter dispersion, then one can observationally assess  the coefficient $C$. 
The brightness of bursts is a boon for measuring dispersive matter effects, but
their varied spectra  \cite{Gruber:2014iza} and  lower frequency afterglows \cite{Meszaros:1996sv,Cenko:2010cg,Zaninoni:2013hca} are a significant confounding factor.  
Because dispersion measurements rely on temporal knowledge of the emission spectra, a single GRB event cannot yet be used to constrain DM properties.  But with a large number of GRB observations located at a variety of redshifts, it is expected that, statistically, random variations amongst the sources should wash out, and the expected redshift and energy dependence that indicate dispersion should survive.  The Fermi Gamma-Ray Space Telescope \cite{fermi_lat} is  dedicated to the task of gamma-ray observations, so that in the future, a sufficient number of GRBs may be used to provide meaningful constraints. Regardless, from our computations of the refractive index for several neutral pointlike DM models \cite{dm_n},  we expect the DM dispersion, i.e., the coefficient $C$, to be extremely small; as a result, any time lags would be immeasurable.     But, the theoretical landscape is rife with DM candidates, and the work in Ref.~\cite{dm_n} only considered a small subset.  Here, we expound upon our previous work by computing the optical dispersion of composite DM comprised of millicharge constituents.

Assessing the size of DM dispersive effects is crucial to evaluating potential claims of beyond-SM physics because matter dispersion is not the only mechanism by which one can achieve such energy-dependent photon time lags.  In theories with Lorentz invariance violation (LIV), the photon's dispersion relation is modified \cite{myers-pospelov,kosto,kost_mewes_em,kost_mewes_grb}.  
Following Ref.~\cite{jacobpiran}, LIV effects can modify the dispersion relation for photons at an energy scale $E_\text{LIV}$
\begin{equation}
E^2 - p^2 = \pm p^2 \left( \frac{p}{E_\text{LIV}} \right)^n \label{disp_rel}
\end{equation}
for some integer $n\ge 1$.  If the modification in the dispersion relation in Eq.~(\ref{disp_rel}) comes with the minus sign (rather than the plus sign), then high energy photons will lag simultaneously emitted lower energy photons.  Photon time lags due to matter dispersion have a different baseline dependence relative to LIV effects   (i.e., they share a different dependence upon the source's redshift $z$), but if $n=2$ in Eq.~(\ref{disp_rel}), the two effects have a common photon energy dependence.  For this reason, it is useful to know the relative magnitudes of the LIV and matter effects at a given baseline so that matter effects cannot confound potential claims of LIV gleaned from GRB photon arrival times.   

Because DM comprises the bulk of the matter in the universe, we will consider its impact on dispersion in detail.  In Ref.~\cite{dm_n}, we found that the potential LIV effects would dwarf any dispersion due to various models of pointlike DM, assuming $E_\text{LIV}$ is around the Planck scale.  In fact, we found that dispersion due to matter effects was irrelevant until energies around $10^{29}$ GeV.  But composite DM models \cite{Faraggi:2000pv, Gudnason:2006ug, Gudnason:2006yj, feng_hiddendm, kaplan_da1, Hamaguchi:2009db, Barbieri:2010mn, Lisanti:2009am, cidm_alves, quirky, formfactor_dm, kaplan_da2, Hur:2007uz,  Blennow:2010qp, 
DelNobile:2011je, Frigerio:2012uc, Cline:2012bz, cline_da, Kumar:2011iy, kouvaris, wallemacq, Holthausen:2013ota,  Buckley:2012ky,    Higaki:2013vuv, cline_strong, Cline:2014eaa, 
Bhattacharya:2013kma, Boddy:2014yra,bro, Carmona:2015haa, Choquette:2015mca,  Wallemacq:2014sta, Antipin:2015xia} 
might prove to be more reactive, particularly if the DM is comprised of charged constituents \cite{quirky, Lisanti:2009am, DelNobile:2011je, Buckley:2012ky, kouvaris, Cline:2012bz, cline_strong, Choquette:2015mca, Cline:2014eaa, Gudnason:2006ug, Gudnason:2006yj, Wallemacq:2014sta, wallemacq, cline_da}.

The motivation for composite dark matter models is varied.  Some models are constructed so as to explain possible photon signals of indirect dark matter detection \cite{kouvaris, Cline:2012bz, Cline:2014eaa,bro}.  Others introduce composite systems designed to smooth out the cuspiness of simulated DM galactic halo profiles \cite{cline_strong,Boddy:2014yra}.
Furthermore, several models attempt to rectify seemingly contradictory results in the experimental search for DM.
 The DAMA/LIBRA experiment \cite{dama_libra} reports a statistically significant annual modulation in its detector which could be attributed to the relative motion of the detector through the galaxy's dark matter halo. Furthermore, the CoGeNT  experiment \cite{cogent1} reports  signals above background in its detector which, if due to dark matter, would be consistent with the apparent signal from DAMA/LIBRA.  If these results are due to DM interactions, they occupy a region of parameter space that has been seemingly ruled out by the CDMS-II \cite{cdmsii_ge} and XENON100 \cite{xe100} experiments.   To reconcile results from DAMA with null results from other experiments, the notion that DM could interact through inelastic channels has been proposed \cite{idm_tuckersmith,idm_chang,idm_marchrussell}.  
A natural way to incorporate inelastic interactions into a model is to allow DM to be composite, rather than point-like, and a host of models take this tack as means to accommodate the DAMA or CoGeNT results in light of other DM constraints
\cite{kaplan_da1,cidm_alves,formfactor_dm,kaplan_da2,cline_da,kouvaris,wallemacq,Lisanti:2009am, DelNobile:2011je,Kumar:2011iy, Blennow:2010qp, Wallemacq:2014sta}.

In what follows, we will consider composite dark matter particles which are electrically neutral, but comprised of millicharged constituents. Taken {\em en masse}, we are interested in the dispersive refractive index of such a medium.  As the DM is electrically neutral, it is essentially invisible to low energy photons.  However, for photons which are near the threshold energy needed to transition the composite DM to an excited state, the photon interaction is substantive.
Considering dark matter as a bulk medium, the interaction between dark matter and light can be characterized in terms of an electric susceptibility and index of refraction.   For photon energies below the transition energy $\omega \ll \omega_0$, the medium will rather generically exhibit dispersion quadratic in the photon energy $n(\omega) \approx 1 + B + C\omega^2$ for constants $B$ and $C$ as with the pointlike DM, because the low-energy theorems of Compton scattering can be generalized to composite structures \cite{brodsky}.

\section{Index of refraction}

Classically, a linear dielectric medium, such as  a dilute gas of dark matter, will acquire a polarization when subjected to an external electric field.  The degree of polarization, or dipole moment per unit volume, can be characterized through the electric susceptibility $\mathbf{P} = \chi_e \mathbf{E}$.  From the susceptibility, we can compute the medium's index of refraction $n =\sqrt{1+\chi_e}$.  To compute the susceptibility of a dark-matter medium, we take a semiclassical approach in which the quantum mechanical DM system interacts with a classical electromagnetic wave via electric dipole transitions.  The constituents which comprise the DM will be assumed to be effectively non-relativistic so that they can be described via the Schr\"odinger equation. We assume the system consists of two constituents of masses $m_{a}, m_{b}$ with electric millicharge $\pm \epsilon e$ bound via a potential $V(r)$ with $r$ the relative separation between the particles.  Though we assume a pair of constituent particles, the analysis can be extended to bound states consisting of more particles if need be. Defining the reduced mass $m := m_a m_b/(m_a + m_b)$ and relative momentum  $\mathbf{p} := m\dot{\mathbf{r}}=(m_b \mathbf{p}_a-m_a \mathbf{p}_b)/(m_a +m_b)$, we construct the unperturbed Hamiltonian $H_0 := - \nabla^2/(2m) + V(r)$ for the system.  
To determine the interaction between the DM and light, a classical electromagnetic wave of frequency $\omega$ interacts with the quantum mechanical electric dipole moment of the DM, $\pmb{\mathpzc{p}} := -\epsilon e \mathbf{r}$, which introduces to the Hamiltonian a perturbation, $H' = -  \pmb{\mathpzc{p}} \cdot \mathbf{E}$.
 In the long wavelength limit, the spatial variation of the electric field is irrelevant leaving the time-dependent perturbation $H' =-\pmb{\mathpzc{p}} \cdot \mathbf{E}_0 \cos \omega t$.

\subsection{Millicharged atomic dark matter}

To create composite particles in a dark sector, modelers introduce a new dark gauge group which results in a binding force among the composite's constituents.  The simplest gauge group is $U(1)$ \cite{feng_hiddendm,kaplan_da1,formfactor_dm,kaplan_da2,cline_da,wallemacq,cline_strong, Cline:2014eaa,Wallemacq:2014sta}.  
If the symmetry is unbroken, then the dark photon is massless and can kinetically mix with the Standard Model (SM) photon.  In such models, the particles can effectively couple to the SM photon thereby acquiring a fractional electric charge  \cite{holdom}.  This permits the existence of dark atoms which are overall electrically neutral, but made of constituents with electric millicharge \cite{cline_da,wallemacq,cline_strong,Cline:2014eaa,Wallemacq:2014sta}.  A neutral dark atom comprised of millicharged particles consists of two fermions $\psi_\mathbf{p}$ and $\psi_\mathbf{e}$, charged under the unbroken gauge group $U(1)'$, coupling to the dark photon with opposite charges $\pm \mathbf{e}$.  [NB: The boldface type is meant to refer to the particles and couplings in the dark sector.]  The dark ``proton" and ``electron" can form bound states under the dark Coulombic potential $V(r) =  -\boldsymbol{\alpha}/r$, where we define the dark fine structure constant $\boldsymbol{\alpha} := \mathbf{e}^2/ (4\pi)$, and the relative separation between the particles is $r$.   We can use non-relativistic quantum mechanics to describe this dark ``hydrogen", $\mathbf{H}$.  Without loss of generality, we assume $m_\mathbf{e}\le m_\mathbf{p}$ and define the reduced mass and relative momentum as above with $m_a= m_\mathbf{e}$ and $m_b= m_\mathbf{p}$.  The unperturbed Hamiltonian $H_0$ yields the usual  hydrogenic eigenstates $\psi_{n\ell m}$ and energy spectrum $E_n = - \boldsymbol{\alpha}^2 m/(2n^2)$ indexed by principal quantum number $n$, a positive integer. The dark and SM sectors are coupled through photon kinetic mixing which gives rise to electric millicharges $\pm \epsilon e$ of the dark particles.

Electromagnetic waves can induce transitions  between dark atomic energy states, but we argue that the bulk of the dark atoms exist in the ground state.  There are three main mechanisms by which the atoms can be excited beyond the ground state: dark atom self-interactions, absorption of dark photons,  or absorption of SM photons.    The existence of elliptical DM halos severely constrains the DM self-interaction cross section; from the limits in Ref.~\cite{MiraldaEscude:2000qt}, the ratio of the DM self interaction to the dark atom's mass must be $\sigma/m_\mathbf{H} < 0.02$ cm$^2$/g though more recent studies have relaxed this bound to 0.1 cm$^2$/g  \cite{Peter:2012jh}.
These limits can be satisfied
either through tuning the model parameters or dilution of the dark atom component of DM.   For models which satisfy this constraint, we can estimate the mean free time between collisions for dark atoms in the Milky Way's galactic halo.  The mean free path can be estimated as $\lambda \sim (\sigma N)^{-1}$ where $N$ the number density of dark atoms; the number density is related to the DM mass density via $N = \rho/m_\mathbf{H}$.  Then the time between collisions is $t_\text{fp} \sim \lambda/v =  m_\mathbf{H} /(\sigma \rho v) $.  Taking as typical parameters the local dark matter density $\rho \sim 0.3$ GeV/cm$^3$ and $v\sim 200$ km/s, we find $t_\text{fp} \gtrsim  10^{18}$ s; i.e., they are non-interacting. 
Dark-atom absorption of dark photons also produces excited states.  The greatest energy density of dark radiation is found in the dark analog of the cosmic microwave background (CMB).  Viable models require the dark radiation to be slightly cooler than the CMB  \cite{adm_cosmo}, so these dark photons will not have sufficient energy to excite the dark atoms.  All that remains is the interaction with SM electromagnetic waves, which we discuss below.

We consider a dark atom in its ground state which can be excited by SM photons.  We  restrict our study to a two-state system, limiting the electromagnetic wave frequency to $\omega \lesssim \omega_0 := E_2-E_1$.  In the presence of the electromagnetic wave, the Hamiltonian for the dark atom is $H= H_0 + H'$, and a general state is $\Psi(t) = c_1(t) e^{-i E_1t} \psi_1+c_2(t) e^{-i E_2t} \psi_2$, with stationary eigenstates $\psi_{1,2}$.  Using the Schr\"odinger equation, we can develop coupled differential equations for the coefficients $c_1$ and $c_2$.  These equations must be amended to account for spontaneous emission of a dark photon from the excited state.  With this extra term proportional to the decay constant $\Gamma$, we have the equation for $c_2$
\begin{equation}
\frac{\mathrm{d}}{\mathrm{d}t} c_2(t) = -i \, (\pmb{\mathpzc{p}}_{21}\cdot \mathbf{E}_0) e^{i\omega_0t} \cos(\omega t) c_1 (t)   -\Gamma c_2(t)
\end{equation}
where $ \pmb{\mathpzc{p}}_{21} := \langle \psi_2 | \pmb{\mathpzc{p}}  | \psi_1  \rangle$ and $\Gamma=  \frac{2^7}{3^8}\boldsymbol{\alpha}^5 m$, adapted from atomic hydrogen \cite{loudon}.

We solve perturbatively for the coefficients $c_{j}(t) = c_{j}^{(0)}(t)+ c_{j}^{(1)}(t)+\cdots$  with initial conditions $c_1(0)=1$ and $c_2(0)=0$.  To determine the {\em linear} response of the atom to the field $E_0$, we only need the zeroth order approximation for $c_1 \approx 1$ and the first order approximation for $c_2$
\begin{equation}
c_2(t) \approx -\frac{1}{2} (\pmb{\mathpzc{p}}_{21}\cdot \mathbf{E}_0) \left[ \frac{e^{i(\omega_0 + \omega)t}}{\omega_0 + \omega - i \Gamma}+\frac{e^{i(\omega_0 - \omega)t}}{\omega_0 - \omega - i \Gamma }\right].
\end{equation}
The induced dipole moment for $\Psi$ is thus
\begin{equation}
\pmb{\mathpzc{p}}(t) = -\epsilon e \langle \Psi(t) | \mathbf{r} | \Psi(t) \rangle = -2\epsilon e\,  \mathrm{Re} [ \mathbf{r}_{12} c_1^*c_2 e^{-i \omega_0 t}].
\end{equation}
We note that the polarization $P(t)$ will be the product of the average induced dipole moment in the direction of $\mathbf{E}_0$ and the number density $N$.  Averaging over the relative orientation between $\mathbf{r}_{12}$ and $\mathbf{E}_0$,   the susceptibility is
\begin{equation}
\chi_e = N\pi\frac{2^{18} }{3^{11}}  \frac{\epsilon^2 \alpha }{\boldsymbol{\alpha}^2 m^2}\frac{\omega_0}{\omega_0^2 -(\omega+i\Gamma)^2}.  
\end{equation}

Given that the DM medium is weakly interacting and dilute, $\chi_e$ is nearly zero.  As such we can approximate the index of refraction as $n \approx 1 +\frac{1}{2} \chi_e$.  For frequencies below the transition energy $\omega < \omega_0$, the index of refraction exhibits normal dispersion; that is, $n(\omega)$ increases with $\omega$.  In fact, for $\omega \ll \omega_0$, the dispersion is quadratic in frequency
\begin{equation}
\mathrm{Re}(n) \approx 1  +  N\pi\frac{2^{20} }{3^{12}}  \frac{\epsilon^2 \alpha }{\boldsymbol{\alpha}^4 m^3}\left( 1 + \frac{\omega^2}{\omega_0^2}\right) , \label{adm_index}
\end{equation}
neglecting the small term proportional to $\Gamma^2$.

\subsection{Other millicharged composite particles \label{sho_sect}}

Millicharged atomic DM represents only a fraction of proposed composite DM models.  Many models posit composite states strongly bound by a nonabelian gauge force \cite{Holthausen:2013ota, quirky, Lisanti:2009am,DelNobile:2011je,Buckley:2012ky,kouvaris,Cline:2012bz,cline_strong,Choquette:2015mca,Barbieri:2010mn,Hur:2007uz,Faraggi:2000pv,Kumar:2011iy,Blennow:2010qp,Hamaguchi:2009db,Higaki:2013vuv, Bhattacharya:2013kma, Carmona:2015haa,Gudnason:2006ug,Gudnason:2006yj,Antipin:2015xia,Boddy:2014yra,bro}, some of which  contain electrically charged constituents \cite{quirky,Lisanti:2009am, DelNobile:2011je,Buckley:2012ky, kouvaris, Cline:2012bz, cline_strong,Choquette:2015mca, Gudnason:2006ug, Gudnason:2006yj}. The details of these strongly composite systems with charged constituents are model dependent, so we will only sketch an approach as to how one would estimate the susceptibility of such a DM medium.  As with the atomic DM system, we will assume a composite state consisting of two particles with effective (dressed) masses of $m_a$ and $m_b$ with electric millicharges $\pm \epsilon e$.  To model a tightly bound system, we approximate their interaction potential as that of a harmonic oscillator,  $V(r) = \frac{1}{2}m\omega_0^2 r^2$ with $m$ the reduced mass. As above, we will only consider a two state system consisting of the ground state and first excited state; the energy difference between these states is $\omega_0$.
Our na\"ive assumptions result in a simplistic model, yet for a system as complex as a nucleon, the SHO potential adapted to three constituent quarks yields an order of magnitude estimate of the nucleon polarizability \cite{holstein}.

With this new potential, we merely need to compute the transition dipole moment $\mathpzc{p}_{21}$ and the decay constant $\Gamma$.  We find $\mathpzc{p}_{21} = \epsilon e/\sqrt{2m \omega_0}$ and $\Gamma =\epsilon^2 \alpha \omega_0^2/(3m)$. Given our assumption of nonrelativistic QM, $\Gamma/\omega_0$ is small so that we can approximate the susceptibility of this DM medium as
\begin{equation}
\chi_e = \frac{4}{3} \pi N \frac{\epsilon^2 \alpha }{ m \omega_0^2}\left( 1 + \frac{\omega^2}{\omega_0^2}\right).  
\end{equation}
Finally, because the susceptibility is small, the refractive index can be approximated by $n \approx 1 + \frac{1}{2} \chi_e$ for $\omega \ll \omega_0$.

\section{Observational consequences for GRB photons}

As a broadband pulse of electromagnetic radiation travels through a dispersive medium, the pulse shape spatially broadens because the phase speed of each component wave is frequency dependent.  The DM medium we consider is normally dispersive, $n \approx 1 + B + C \omega^2$, so higher frequency components of the pulse will lag the lower frequency components.  
Over large distances, a time lag can accrue between these two components.  Gamma-ray bursts are apt photon sources for dispersive studies because they are explosive events that occur over short time scales and their brightness permits observation over cosmological distances. 
Because cosmological distances are involved, we must account for the universe's expansion as a light pulse travels from source to observer \cite{jacobpiran}.  There are three factors that need to be considered.  First, the light-travel time between source and detector is dependent upon the redshift of the source and the local expansion rate.  Second, at redshift $z$ the number density of dark matter increases by a factor of $(1+z)^3$; a factor of present day DM number density  $N$ is contained in the coefficient $C$.  Finally, as we look into the past, the wavelength of light blue shifts relative to its value $\omega$ at the detector (at $z=0$). Incorporating these three factors, the time lag accrued between (detected) frequencies $\omega_\text{hi}$ and $\omega_\text{lo}$ simultaneously emitted from a source at redshift $z$ becomes
\begin{equation}
\tau \approx C\left( \omega_\text{hi}^2 -\omega_\text{lo}^2 \right) \int_0^z \frac{(1+z')^5 \mathrm{d}z'}{H(z')},   \label{matter_lag1}
\end{equation}
where   the Hubble expansion rate at redshift $z'$ is $H(z') = H_0 \sqrt{(1+z')^3\Omega_M + \Omega_\Lambda}$, assuming a simple $\Lambda$CDM cosmology. If the two photon energies are well separated, we can neglect the low energy term in the difference in Eq.~(\ref{matter_lag1}) and set $\omega = \omega_\text{hi} \gg \omega_\text{lo}$ so that
\begin{equation}
\tau \approx C\omega^2 K_5 (z),  \label{matter_lag2}
\end{equation}
where we define the integral over the baseline $K_j (z) := \int_0^z \frac{(1+z')^j \mathrm{d}z'}{H(z')}$.
We employ the cosmological parameters in the 2015 data release from the Planck satellite \cite{planck2015}.
The Hubble constant today is $H_0 = 67.8 \pm 0.9 \,\hbox{km/s/Mpc}$, whereas 
the fraction of the energy density
in matter relative to the critical density today is $\Omega_M = 0.308\pm 0.012$.  For the simple $\Lambda$CDM model, the universe is flat so that  the corresponding fraction of the energy density 
in the cosmological constant $\Lambda$ is 
$\Omega_\Lambda = 1- \Omega_M$.

The high energy photons from distant GRBs have already been used to place lower bounds on the scale at which Lorentz-invariance violating effects could modify the photon's dispersion relation  \cite{AmelinoCamelia:1997gz,ellis, 
Boggs:2003kxa,Ellis:2005wr,jacobpiran, ellis2,Ellis:2009yx,AmelinoCamelia:2009pg,Vasileiou:2013vra}.  For some LIV models, the modified dispersion relation acquires terms quadratic in the photon energy, i.e., $n=2$ in Eq.~(\ref{disp_rel}).   For these models, the following time lag can accrue between the low and high energy photons emitted by a GRB
\begin{equation}
\Delta t_\text{LIV} \approx \frac{3}{2} \left(\frac{\omega}{E_\text{LIV}}\right)^2 K_2(z).  \label{t_liv}
\end{equation}
where the (present day) photon energy $\omega = \omega_\text{hi} \gg \omega_\text{lo} $ was emitted at redshift $z$.  The energy $E_\text{LIV}$ characterizes the scale at which LIV effects become appreciable.  
Data from GRBs and AGNs have placed limits on this scale (for quadratic dependence) to be $E_\text{LIV} > 1.3 \times 10^{11}$  GeV \cite{bolmont}.

We wish to assess the size of dispersive matter effects relative to those attributable to LIV.  From Eqs.~(\ref{matter_lag2}) and (\ref{t_liv}), we will consider the ratio $\tau/\Delta t_\text{LIV}\sim \frac{2}{3} C E_\text{LIV}^2 K_5(z)/K_2(z)$ for millicharged composite DM candidates.  The factor $C$ is dependent on the DM model, whereas the ratio of the $K_j$ integrals depends on the redshift of the source with $K_5(z)/K_2(z)\sim \mathcal{O}((1+z)^3)$.  The GRB constraint in Ref.~\cite{bolmont} is derived from GRB 090510 which is located at a reshift of $z=0.903\pm 0.003$ with a high energy photon $\omega_\text{hi}=30.53^{+5.79}_{-2.56}$ GeV detected 0.829 s after the trigger of the GRB monitor  \cite{fermi_nat}.  In this case, the ratio of integrals $K_j$ is $\mathcal{O}(10)$.  On the other hand, there have been two GRB observations with confirmed redshift $z>8$:  GRB 090423 with $z=8.3\,$ \cite{Tanvir:2009zz} and GRB 090429B with $z=9.4$ \cite{Cucchiara:2011pj}. For these, we find the maximum value of  $K_5/K_2 \sim \mathcal{O}(10^3)$.  Thus, matter effects are enhanced dramatically relative to LIV effects for higher redshift sources.

As stated previously, we work within the context of a simple $\Lambda$CDM model; dark energy is attributed to a cosmological constant with an equation of state $w= -1$, where $w$ is the ratio of DE's pressure to energy density.  This is consistent with the 2015 Planck data which, along with external astrophysical data, determines $w=-1.006 \pm 0.045$, if a constant $w$ is assumed  \cite{planck2015}. 
However, when the Planck data are combined with weak lensing data, cosmologies with a time-dependent equation of state are at least marginally preferred \cite{planck2015de}. 
Extensions of the simple $\Lambda$CDM model will affect the expansion rate of the universe, $H(z')$, and thus the integrals $K_j(z)$.  The impact of cosmology upon GRB photon and neutrino time lags due to LIV was previously considered in Refs.~\cite{Biesiada:2007zzb,Biesiada:2009zz}, where the authors compare results from a simple $\Lambda$CDM model to quintessence, Chaplygin gas, and braneworld cosmologies.  Amongst the models, differences in time lags exist which could affect the measurability of the effect or the interpretation of a measurement if the cosmology is unknown.  Using the models considered in  Refs.~\cite{Biesiada:2007zzb,Biesiada:2009zz}, we determine the impact of cosmology upon the {\em relative} size of dispersive matter effects with those from LIV. 
We find the ratio of integrals $K_5(z)/K_2(z)$ varies little (less than $5\%$) out to $z=10$ for the $\Lambda$CDM, quintessence, Chaplygin gas, and braneworld models.  On the other hand, for the variable quintessence model,  we find that $K_5(z)/K_2(z)$ is commensurate in size with this ratio in the other models up to $z=2$, but beyond this redshift, the ratio of integrals approaches an asymptotic value around 18 (while the ratio scales as $\mathcal{O}((1+z)^3)$ for the other models).    For GRBs at a distant redshift of $z=10$, the difference between the variable quintessence model and the others will be a factor of 25.  On the face of it, this difference is substantial and could perhaps be even larger for different models of quintessence; however, when we examine the size of DM dispersive effects below, we will see that the choice of cosmology is of subleading significance. Given this, we
opt to use the simple $\Lambda$CDM model, though acknowledge that other cosmologies are likely and will quantitatively affect our results to a degree.

\subsection{Millicharged atomic dark matter}

We now examine the size of the dispersive coefficient $C$ in the refractive index  relative to $E_\text{LIV}$.
This coefficient  sets the energy scale, $C^{-\frac{1}{2}}$,  at which DM dispersion becomes appreciable.   For atomic dark matter, we find from Eq.~(\ref{adm_index}) this coefficient to be $C= N\pi\frac{2^{20} }{3^{12}}  \frac{\epsilon^2 \alpha }{\boldsymbol{\alpha}^4 m^3 \omega_0^2}$.   The DM number density is $N = \rho/m_\mathbf{H}$ where the mass of the dark atom is $m_\mathbf{H} = m_\mathbf{p} + m_\mathbf{e} -\frac{1}{2}\boldsymbol{\alpha}^2 m$.  For propagation of light across cosmological distances, the average DM density is $\rho \simeq 1.25\times 10^{-6}$ GeV/cm$^3$\cite{planck2015}.  As a benchmark, it is useful to estimate the value of the coefficient for various atomic dark matter models.  In Ref.~\cite{wallemacq}, the best-fit model employs a heavy dark proton $m_\mathbf{p} = 650$ GeV and much lighter dark electron $m_\mathbf{e}= 0.426$ MeV with an electric millicharge $\epsilon = 6.7\times 10^{-5}$ and a dark fine structure constant which takes the SM value, $\boldsymbol{\alpha} = \alpha$.  For this model, the dispersive term is $C = 1.8 \times 10^{-25}$ GeV$^{-2}$.  As a consequence, the energy scale at which the DM dispersion is appreciable is $C^{-\frac{1}{2}} = 2.4 \times 10^{12}$ GeV, which is commensurate with the limit on $E_\text{LIV}$ from Ref.~\cite{bolmont}.  On the face of it, it would seem that for large $z$ the $K_5$ integral could make dispersion due to millicharge atomic DM competitive in magnitude with potential LIV dispersive effects, but we must recall one crucial point--the binding energy of the dark atom.  Our estimate of the DM refractive index is only valid in the limit in which observed photon energies are much smaller than the energy difference between the ground and excited state of the dark atom, $\omega \ll \omega_0 = \frac{3}{8} m \boldsymbol{\alpha}^2$.  The dark electron in Ref.~\cite{wallemacq} has a mass similar to the SM electron, so the ionization energy of the dark hydrogen is on the scale of electronvolts with $\omega_0 = 8.5$ eV.  As such, the limits derived from high energy GRB photons are not relevant for this model of atomic dark matter because the dark electrons are so weakly bound.  Let us now  consider another model of atomic dark matter which employs very different masses and electric millicharge.  Reference \cite{cline_da} has an atomic DM model which can account for the CoGeNT experimental results. The authors' dark atom consists of massive particles $m_\mathbf{p} =   m_\mathbf{e} = 3$ GeV with a slightly larger dark fine structure constant $\boldsymbol{\alpha}= 0.062$ and much larger millicharge $\epsilon = 10^{-2}$.  In this case, the relevant energy scale for matter dispersive effects is $C^{-\frac{1}{2}} = 6 \times10^{21}$ GeV with transition threshold of $\omega_0 = 2.2$ MeV.  Due to the larger masses and dark fine structure constant, the threshold energy $\omega_0$ rises, allowing the dark matter to be probed with more energetic photons, but for such massive models, the energy scale at which matter effects become appreciable is well beyond the current bounds on $E_\text{LIV}$.

\begin{figure}
\includegraphics[width=10cm]{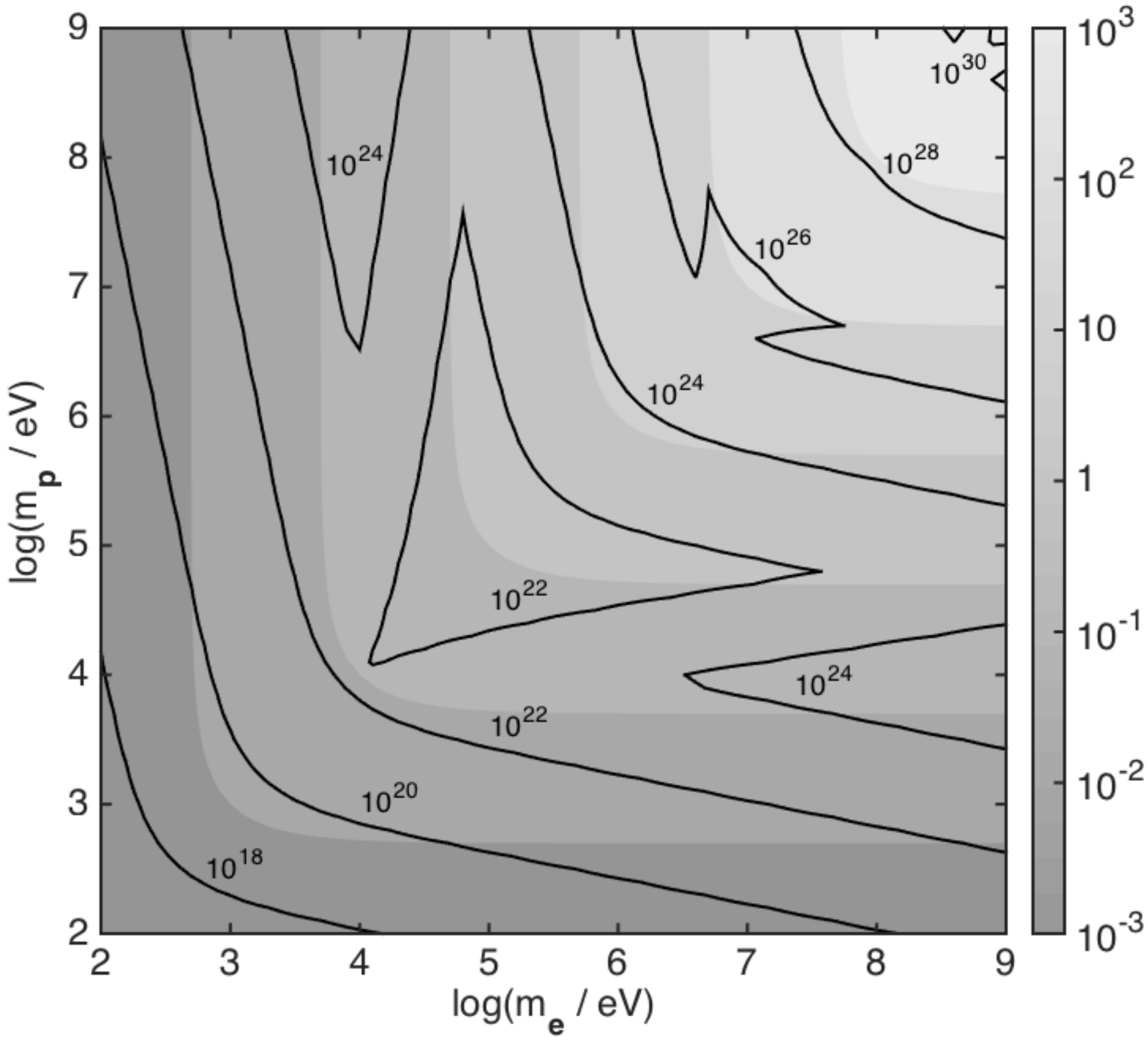}
\caption{The solid  contour lines depict $C^{-\frac{1}{2}}$ for millicharged atomic dark matter; this is superimposed upon  filled contours which represent the threshold energy $\omega_0$.
The dark atom consists of a dark electron and proton with mass $m_{\mathbf e}$ and $m_{\mathbf p}$ and dark fine structure constant equal to that of the SM, $\boldsymbol{\alpha} = \alpha$. All contours carry units of electronvolts. \label{fig1}}
\end{figure}

These two specific examples are exemplars of a general trend.  Namely, for light DM masses, the energy scale at which matter dispersion is operative is commensurate with limits on the LIV energy scale, but the binding energy is too small to actually probe such dispersion. For more massive models, the threshold energy $\omega_0$ increases, but $C^{-\frac{1}{2}}$ increases at a greater rate.  To confirm these generalities, we explore more fully the allowed parameter space of exotic millicharged dark atoms without regard to their feasibility as a dark matter candidate.
In particular, we compute the dispersive energy scale $C^{-\frac{1}{2}}$ for a range of dark electron and proton masses subject only to the provision that the existence of the constituent millicharged particles have not been excluded through other considerations.  There are, in fact, stringent constraints on electric millicharge \cite{davidson,Vogel:2013raa}.  Figure 1 of Ref.~\cite{Vogel:2013raa} shows a current summary of these constraints for particle masses ranging from 100 eV to 100 TeV.  Stellar evolution severely constrains electric millicharge $\epsilon < 2\times 10^{-14}$ for masses below 10 keV.  Around 100 keV to 1 GeV, BBN and CMB constraints upon the light degrees of freedom limit electric millicharge $\epsilon \lesssim 10^{-9}$.  For masses between 1 MeV and 100 GeV,  collider constraints limit $\epsilon < 0.2$.
Dispersive effects scale as $\epsilon^2$, so they will be maximized for the largest allowed $\epsilon$. To explore the largest possible matter dispersion, we choose, for a given particle mass, the millicharge  which saturates the bounds in Fig.~1 of Ref.~\cite{Vogel:2013raa}.  Setting the dark fine structure constant equal to the SM value $\boldsymbol{\alpha} = \alpha$, we plot in Fig.~\ref{fig1} contours representing the  energy scale $C^{-\frac{1}{2}}$ for various dark electron and proton masses.  These contours are superimposed upon filled contours which represent the threshold energy $\omega_0$.  For masses ranging from 100 eV to 1 GeV, we find the energy scale charactering matter dispersion range from $10^9$ GeV to $10^{21}$ GeV, spanning twelve decades of energy, while $\omega_0$ ranges from meV to keV, spanning six decades. 

The broad scan of parameter space depicted in Fig.~\ref{fig1} confirms our preliminary conjecture. We see rather generally that if the energy scale at which atomic DM dispersive effects are near current limits on $E_\text{LIV}$ then the binding energy of the dark atom is sub-eV.  The photon energy needed to probe the DM dispersion would ionize the dark atoms. For higher threshold energies, matter dispersive effects quickly become irrelevant.  Supposing a LIV scale around the Planck mass, $\sim 10^{19}$ GeV, the ionization energy of the dark atom is on the order of keV.
Thus, in a search for LIV dispersive effects, dispersive effects due to atomic dark matter are not a confounding source of background.   
 
\subsection{Other millicharged composite particles}

We now turn to our millicharged composite system bound under the harmonic oscillator potential. There is wide variation amongst the strongly bound composite DM models, but we will explore the dispersive matters effects for a limited range of particle mass and oscillator energy $\omega_0$.  Again, to determine the maximal effect, we assume electric millicharge values which saturate the bounds in Fig.~1 of Ref.~\cite{Vogel:2013raa}.  From above, we find the dispersive coefficent to be  $C  =  \frac{2}{3} \pi N \frac{\epsilon^2 \alpha }{ m \omega_0^4}$.  We will assume that the constituent millicharged particles have the same effective mass $m_a = m_b$ so that the reduced mass is one-half this mass $m = \frac{1}{2} m_a$ and the total mass of the system is roughly $2m_a$. With these simplifying assumptions, the dispersive energy scale depends rather simply upon the composite parameters, $C^{-\frac{1}{2}} \sim m_a \omega_0^2/\epsilon$.  As a result, this energy  is lowest for masses $m_a$ in the GeV to 100 GeV region where $\epsilon \lesssim 0.2$.  

\begin{figure}[h]
\includegraphics[width=10cm]{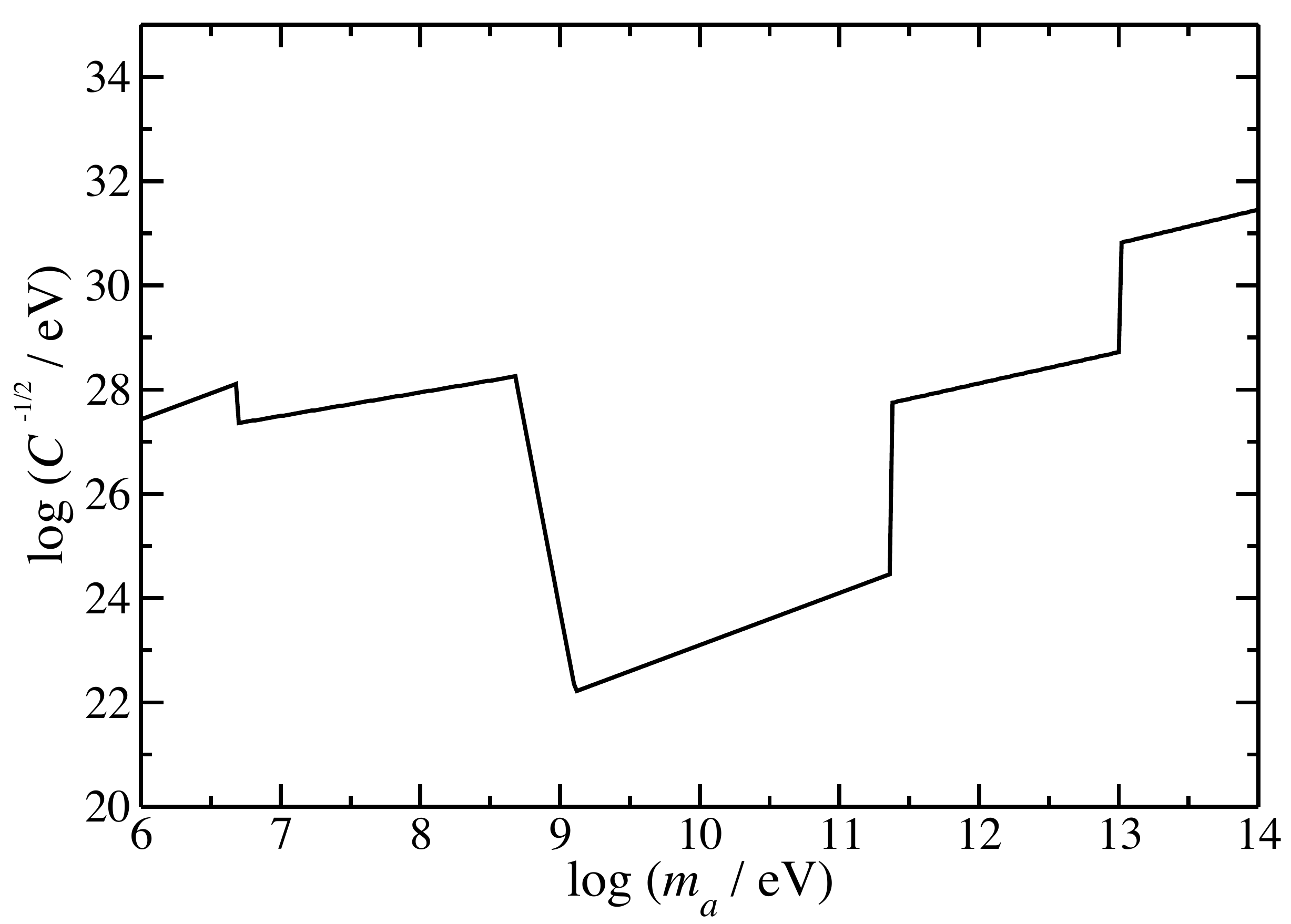}
\caption{ Plot of $C^{-\frac{1}{2}}$ as a function of constituent mass $m_a$.  We assume $m_a = m_b$ and a threshold energy of $\omega_0 = 1$ keV. \label{fig2}}
\end{figure}

In Fig.~\ref{fig2}, we plot the dispersive energy scale as a function of the mass $m_a$, choosing $\omega_0 =1$ keV.  The threshold energy could take a host of values, but we choose 1 keV because  inelastic DM models with an excitation scale of order $\mathcal{O}(1-100$ keV) can accommodate the DAMA anomalous experimental results or potential indirect detection signals \cite{Lisanti:2009am,Cline:2014eaa, Kumar:2011iy, bro}. Given that $C^{-\frac{1}{2}}$ is proportional to $\omega_0^2$, it is not difficult to extrapolate these results to different threshold values.  
Referring to Fig.~\ref{fig2},  we find that the lowest dispersive energy scale, $C^{-\frac{1}{2}}= 1.7 \times 10^{13}$ GeV, occurs for a millicharged particle mass of 1.3 GeV.   This energy is a few decades above the current limit on the LIV scale, but over long baselines the size of matter dispersive effects could rival those due to LIV.  Of course, unlike LIV, the quadratic dispersive terms is only relevant for photon energies below the threshold of 1 keV.  This difference between the threshold and dispersive energy scales makes the matter dispersive effects immeasurable. Indeed, for a nearby GRB, $z\sim 1$, the time lag between  keV and lower energy photons is on the order of $10^{-20}$ s.  Given the same millicharged particle mass of 1.3 GeV, a dispersive scale near the Planck mass, $C^{-\frac{1}{2}} \sim 10^{19}$ GeV, involves a threshold energy near 1 MeV, but probing the DM with near threshold photons results in the same time lag for the $\omega_0 = 1$ keV case. 
As with atomic dark matter, whenever the energy scale of the dispersive coefficient is commensurate with the LIV scale, the threshold energy of the composite system is too small to probe this scale.  Though one has some freedom to tune the threshold of the SHO model independent of the millicharged particle properties, this freedom is not sufficient to construct a composite DM candidate that can have measurable dispersive effects competitive with potential LIV effects.

\subsection{Baryonic matter}

One potential source of background to the dispersive matter effects from composite millicharge DM is ordinary baryonic matter which, for the most part, consists of the hydrogen and helium produced in big-bang nucleosynthesis. Dispersion due to baryonic matter occurs on varying scales, the first being the atomic scale, which is relevant for photon energies in the eV range.  As we discussed above, the probes of LIV are high-energy (MeV or greater) photons that would ionize the hydrogen and helium of baryonic matter, so these cannot be confounded with LIV effects.  Beyond the atomic energy scale, a medium of hydrogen and helium will effectively appear as a plasma to high energy photons, and the plasma's optical properties will be predominantly determined by its electron component. Such a medium possesses a dispersive index of refraction of the form $n \approx = 1- A\omega^{-2}$ with $A= e^2 N/(2m_e)$ where $N$ is the number density of the electrons \cite{cosmicn,dm_n}.  This does not  have the $\mathcal{O}(\omega^2)$ dispersive behavior typical for a neutral scatterer, so it cannot be confused with dispersion due to DM or LIV.  Furthermore, for a low density cloud of atoms, dispersion is negligible for high energy photons because $n -1 \sim 1/\omega^2$.  

For photon energies approaching the nuclear scale, there are $\mathcal{O}(\omega^2)$ dispersive effects to consider arising from the polarizabilities of the nucleons, but this physics is reasonably well understood both from an experimental and theoretical standpoint \cite{holstein_rev}.  As a result, the impact of baryonic dispersion can be studied in detail, but here we opt to estimate the effect by combining measured values of a nucleon's electric polarizability with our SHO model for strongly bound systems in Sec.~\ref{sho_sect}.  From Ref.~\cite{holstein_rev}, we find electric polarizabilities for the proton and neutron, $\alpha_\text{E}^p=11 \times 10^{-4}$ fm$^3$ and $\alpha_\text{E}^n=12 \times 10^{-4}$ fm$^3$, so for a nucleon we take $\alpha_\text{E}^{N}=12 \times 10^{-4}$ fm$^3$.  The small size of the polarizability indicates a tightly bound system, and we expect dispersive effects to be small.
With our SHO model for dispersion, we find $n \approx 1 + 2\pi N \alpha_\text{E}^N(1 + \omega^2/\omega_0^2)$ where $N$ is the number density of nucleons; we estimate the excitation energy to be $\omega_0 \approx 300$ MeV \cite{holstein}.  From the Planck 2015 data, the baryonic component of the universes's energy budget is $\Omega_b =0.048$ so that number density of nucleons is $N = 2.5 \times 10^{-7}$ cm$^{-3}$.  This renders a dispersive term of  $n -1 \approx   (2 \times 10^{-48})\, \omega^2/\omega_0^2$.  For photon energies well below $\omega_0$, the dispersive energy scale is $C^{-\frac{1}{2}} \sim 2\times 10^{26}$ MeV.  In short, this is not competitive with LIV probes and likely immeasurable.  But, even if a suspected baryonic signal arose from a photon time lag, one can assess the presence of hydrogen and helium along the line of sight  through the absorption spectrum that will arise in the optical afterglow of the burst.  This absorption spectrum can aid in determining the presence of atomic scale physics, e.g., a Lyman-alpha system, differentiating the matter from millicharged DM.

\section{Conclusion}

Models of composite dark matter represent an attractive alternative to simple pointlike DM candidates. Through their inelastic interactions, one can account for potential  DM direct and indirect detection signals, produce more realistic DM galactic halo profiles, and provide a natural explanation for the relative abundance of dark to baryonic matter.  In addition to these physical motivations,  composite systems are aesthetically pleasing because the dark sector  mirrors  some of the complexity of the SM sector.  In this work, we focused upon composite systems which consist of electrically charged constituents.  Such particles naturally couple to the electromagnetic field via an electric dipole moment, rendering the cosmos with a dispersive optical index of refraction.  We computed this index of refraction for atomic DM and more strongly bound composite systems, modeled through a harmonic oscillator potential. Given the dispersive nature of the refractive index, higher energy photons will lag those with lower energy as they travel through the universe from a distant GRB.  A time lag accrued over long baselines could be confused with similar dispersive effects which result from theories of LIV. However, we found that, for both atomic and more strongly bound composite DM, whenever the energy scale of matter dispersion is commensurate with that of LIV dispersion, the threshold energy $\omega_0$ is too small to actually probe the matter dispersive effects.  As a result, potential claims of LIV achieved through time lags cannot be confused for matter dispersion due to composite DM comprised of charged constituents.

\section{ACKNOWLEDGMENTS}

AK acknowledges partial support during the completion of this work from the  Washington NASA Space Grant Consortium and  the University of Puget Sound.  DL thanks Bernie Bates for useful discussions. 

\bibliography{biblio}

\end{document}